\documentclass[preprint,aps,
showpacs
]{revtex4}
\usepackage{epsfig}
\begin{document}
%
\preprint{}
\date{\today
}
\title{
S-branes from unbalanced black diholes
}
\author{Kenji~Sakamoto}
\email{sakamoto@het.phys.sci.osaka-u.ac.jp}
\affiliation{Department of Physics, \\
Osaka University, \\
Toyonaka, Osaka 560-0043, Japan}

\begin{abstract}
We construct new non-singular and time-dependent solutions 
from the black diholes with unbalanced magnetic charge. 
These solutions are constructed by the double Wick rotation 
with the analytic continuation of the mass or 
NUT-parameter of unbalanced black diholes. 
In the limit of balanced magnetic charge, our solutions reduce to the S-brane solution obtained from the 
black diholes discussed by Jones et al. \cite{Jones:2004rg}.
We study the behaviors of metric components and 
discuss the s-charge over a constant time-slice. 
From the properties of the solutions, we find that our solutions 
correspond to the S-brane type solutions. 
\end{abstract}
\maketitle


\section{Introduction}
The solutions for multi-black holes have been of great interest. 
The static maximally charged multi-black hole solution was discussed 
by Majumdar and Papapetrou \cite{Papapetrou:1947ib, Majumdar:eu, Myers:rx} and 
other multi-black hole solutions were studied in \cite{IsraelKhan, KS3, Kastor:1992nn, Tan:2003jz}. 
Recently, the static and axisymmetric solutions describing multiple collinear black holes have attracted much attention. 
The multiple collinear Schwarzschild solution was given by Israel and Khan \cite{IsraelKhan}. 
Emparan and Teo considered the static pairs of oppositely charged extremal black holes (black diholes) solution \cite{Emparan:1999au,Emparan:2001bb,Teo:2003ug}. 
These black hole solutions enable us to study their thermodynamics and the interaction between black holes \cite{Emparan:2001bb, Costa:2000kf}. 

Recently, the studies of the S-brane solution have received much attention 
\cite{Gutperle:2002ai, Chen:2002yq, Ohta:2003uw, Kruczenski:2002ap}. 
The S-brane solutions are constructed by the analytic continuation of black hole solutions 
\cite{Burgess:2002vu, Tasinato:2004dy, Astefanesei:2005eq, Jones:2004pz}. 
In the case of multi-black holes, the black dihole solutions lead to non-singular S-brane type solutions 
by double Wick rotation \cite{Jones:2004rg, Biswas:2004zc, Jones:2005hj}. 
S-brane solutions are time-dependent gravitational configurations, 
describing a shell of radiation coming in from infinity and creating an unstable brane which subsequently decays. 
Such solutions are very interesting, because they are singularity-free and periodic in imaginary time.  
S-brane obtained from the collinear black holes solution provide 
large $N$ duals of unstable D-brane in string theory \cite{Jones:2004rg}. 

In this work, we consider the non-singular and time-dependent solutions introduced with double Wick rotating multi-black holes. 
Jones et al. discussed the S-brane solutions constructed from the black diholes which have the same magnitude and opposite magnetic charge \cite{Jones:2004rg}. In our study, we consider the double Wick rotation of the black dihole solutions with the unbalanced magnetic charges, such as the different magnitudes and opposite magnetic charges. 
From the studies of S-brane constructed from the unbalanced black diholes, 
we expect that we can discuss the unstable D-brane with the different magnitudes of charges. 
The obtained solutions contain the NUT-parameter which represents the total magnitude. 
We investigate the singularity and s-charge of our solutions. 
From their properties, we see that our solutions are corresponding to the S-brane type solutions. 
\section{Black Diholes with unbalanced magnetic charges}
The solutions of black diholes with unbalanced magnetic charges are discussed by Liang and Teo \cite{Liang:2001sp}. 
We consider the static, axisymmetric solution of the Einstein-Maxwell-Dilaton 
system. 
\begin{equation}
S=\int d^4x \sqrt{-g} \left( R-2(\nabla\phi)^2 -e^{-2\alpha\phi}F^2 \right),
\end{equation}
where $R$ is the Ricci scalar, $\phi$ the dilaton field and $F_{ab}$ the electromagnetic field tensor. 
We choose a purely magnetic field $A\equiv A_{\varphi}$ and other components of $A_{a}$ to be zero. 
Then, we obtain the black dihole solution represented by 
\begin{equation}
 ds^2= \Lambda^{\frac{2}{1+\alpha^2}} \left( -dt^2
+ \frac{\Sigma^{\frac{4}{1+\alpha^2}}}{(\Delta+
(m^2+a^2-l^2)\sin^2\theta)^{\frac{3-\alpha^2}{1+\alpha^2}}}
(\frac{dr^2}{\Delta} + d\theta^2) \right) +
\frac{\Delta \sin^2\theta}{\Lambda^{\frac{2}{1+\alpha^2}}} d\varphi^2 , \label{solbl}
\end{equation}
and magnetic field and dilaton field are of the following form: 
\begin{equation}
A_{\varphi} d\varphi=-\frac{2}{\sqrt{1+\alpha^2}} \frac{a(mr-l^2)\sin^2 \theta+l \Delta \cos\theta}{\Delta + a^2 \sin^2\theta} d\varphi,
\end{equation}
\begin{equation}
\phi=-\frac{\alpha}{1+\alpha^2} ln \left(\frac{\Delta+a^2 \sin^2\theta}{\Sigma} \right),
\end{equation}
where 
\begin{equation}
\Delta= r^2-2mr-a^2+l^2 ,\hspace{.5cm} \Sigma= r^2 -(a~\cos\theta+l)^2 , \hspace{.5cm}
\Lambda=\frac{\Delta+a^2~\sin^2\theta}{\Sigma} \ .
\end{equation}
This solution describes a pair of extremal dilatonic black holes with unbalanced charges lying on the symmetry axis. 
The parameter $a$ is a measure of the distance between the two black holes. $l$ is the NUT-parameter, representing the monopole field strength of the solution at far distance. 
The curvature singularities (which represent black holes) are located at $r=r_+\equiv m+\sqrt{m^2+a^2-l^2}$ and $\theta=0,\pi$. 
From the asymptotic behaviors of $g_{tt}$ and $A_{\varphi}$, the total mass 
is found to be $M=2m/(1+\alpha^2)$ and the net magnetic charge is $Q=2l/\sqrt{1+\alpha^2}$. 
In the case of $l=0$ and $\alpha=0$, this solution becomes the black magnetic dihole solution studied by Emparan \cite{Emparan:1999au}. The solution for $a\to\infty$ represents the extremal dilatonic black holes \cite{Gibbons:1987ps}.
 
\section{Weyl form}
We rewrite the black dihole solutions by the Weyl form.
The coordinate transformation between $r$, $\theta$ and Weyl coordinates is
\begin{eqnarray}
\rho&=&\sqrt{r^2-2mr-a^2+l^2}\sin\theta\hspace{.4cm}=\sqrt{\Delta}\sin \theta, \\
z&=&(r-m)\cos \theta.
\end{eqnarray}
In the Weyl form, the solution (\ref{solbl}) are given by 
\begin{equation}
ds^2 = -f dt^2 +\frac{e^{2\gamma}}{f}(d \rho^2 +dz^2) +\frac{\rho^2}{f} d\varphi ^2 \label{weyl1},
\end{equation}
where
\begin{eqnarray}
f&=&\left[ \frac{(R_++R_-)^2-4m^2+4l^2-{a^2\over m^2+a^2-l^2}(R_+-R_-)^2}
{(R_++R_-+2m)^2-\{ {a\over \sqrt{m^2+a^2-l^2}}(R_+-R_-)+2l \}^2 }\right]^\frac{2}{1+\alpha^2} \label{fweyl}, \\
e^{2\gamma}&=&\left[ {(R_++R_-)^2-4m^2+4l^2-{a^2\over
m^2+a^2-l^2}(R_+-R_-)^2
\over 4 R_+ R_-}\right]^\frac{4}{1+\alpha^2}, \\
R_{\pm}&=& \sqrt{\rho^2+(z \pm \sqrt{m^2+a^2-l^2})^2}.
\end{eqnarray}
The magnetic field and dilaton field are 
\begin{eqnarray}
A&=&-\frac{1}{\sqrt{1+\alpha^2} \left[(R_++R_-)^2-4m^2+4l^2-\frac{a^2}{m^2+a^2-l^2}(R_+-R_-)^2 \right] } \nonumber \\
& &\hspace{.5cm}\times\Big[ \left\{ am(R_++R_-+2m)-2 al^2 \right\} (4-\frac{(R_+-R_-)^2}{m^2+a^2-l^2})\nonumber \\
& &\hspace{1cm}+\frac{l(R_+-R_-) }{\sqrt{m^2+a^2-l^2}} \left\{(R_++R_-)^2-4(m^2+a^2-l^2) \right\}\Big] 
~ d \varphi , \hspace*{.6cm}\\
\phi&=&-\frac{\alpha}{1+\alpha^2}ln\left[ \frac{(R_++R_-)^2-4m^2+4l^2-{a^2\over m^2+a^2-l^2}(R_+-R_-)^2}
{(R_++R_-+2m)^2-\{ {a\over \sqrt{m^2+a^2-l^2}}(R_+-R_-)+2l \}^2 }\right] \label{weyl2}. 
\end{eqnarray}
In this frame, the extremal black holes are located at $z=\pm\sqrt{m^2+k^2-l^2}$, $\rho=0$. The horizons of these black holes are degenerate. 
The metric has conical singularities along $z$-axis between 
$-\sqrt{m^2+k^2-l^2}<z<\sqrt{m^2+k^2-l^2}$.
\section{S-dihole solutions}
We consider new non-singular, time-dependent solutions which arise by the double Wick rotating black dihole solutions. 
Jones et al. \cite{Jones:2004rg} studied the non-singular S-brane solution from the black diholes 
with the double Wick rotation,  represented by analytically continuing the coordinates
\begin{equation}
t \to iy \ ,\hspace{.5cm} z\to i\tau. 
\end{equation}
Many gravity solutions can be analytically continued to obtain new time dependent solution and
those are not uncommon to have two or more different analytic continuations \cite{Jones:2004pz}. 
In our case, two types of S-dihole solutions can be constructed. We call them S-dihole I and S-dihole II. 
The S-dihole I is obtained by double Wick rotation and the analytic 
continuation of NUT-parameter $l\to -il$.
For the S-dihole II solution, we consider the analytic continuation of mass parameter $m\to im$ with double Wick rotation 
\footnote{This analytic continuation is similar to the one used 
to find the S-brane solutions in \cite{Wang:2002fd, Burgess:2002vu, Tasinato:2004dy}. 
}. 
Under these analytic continuations, the metric components have a real parameter. 
The difference of the two solutions is in the spacetime structure. 
The spacetime structure of S-dihole II is complicated.
The S-dihole II has the same structure with the S-dihole constructed from the balanced black diholes discussed in \cite{Jones:2004pz}, 
except that the positions of the coordinate singularities in S-dihole II depend on the NUT-parameter. 
\subsection{S-dihole I}
One of the non-singular and time-dependent solutions is constructed by 
the following double Wick rotation and analytic continuation: 
\begin{equation}
t \to iy \ ,\hspace{.5cm} z\to i\tau \hspace{.5cm} l\to -il.
\end{equation}
Then, the metric takes the following form:
\begin{equation}
ds^2 = \frac{1}{f} \left(\rho^2 d\varphi ^2 + e^{2\gamma}(d \rho^2 -d \tau^2) \right) +f dy^2, \label{eq18}
\end{equation}
where
\begin{eqnarray}
R&=& \sqrt{\rho^2-(\tau + i\sqrt{m^2+a^2+l^2})^2},\\
f&=&\left[ \frac{(\mathrm{Re} R)^2-m^2-l^2+{a^2\over m^2+a^2+l^2}(\mathrm{Im} R)^2}
{(\mathrm{Re} R+m)^2+\{ {a\over \sqrt{m^2+a^2+l^2}}(\mathrm{Im} R)+l \}^2 }\right]^\frac{2}{1+\alpha^2},\\
e^{2\gamma}&=&\left[ {(\mathrm{Re} R)^2-m^2-l^2+{a^2\over
m^2+a^2+l^2}(\mathrm{Im} R)^2
\over |R|^2}\right]^\frac{4}{1+\alpha^2}.\\
\end{eqnarray}
The gauge field and dilaton field are written as
\begin{eqnarray}
A&=&\frac{-2}{\sqrt{1+\alpha^2} \left[ (\mathrm{Re} R)^2-m^2-l^2+\frac{a^2}{m^2+a^2+l^2}(\mathrm{Im} R)^2 \right]} \nonumber \\
& &\times \Big[\left\{ am(\mathrm{Re} R+m)+al^2 \right\} (1+\frac{(\mathrm{Im} R)^2}{m^2+a^2+l^2})
-\frac{l(\mathrm{Im} R) \left\{(\mathrm{Re} R)^2-(m^2+a^2+l^2)\right\} }{\sqrt{m^2+a^2+l^2}} \Big] 
~ d \varphi , \hspace*{.6cm}~~
\end{eqnarray}
\begin{eqnarray}
\phi=-\frac{\alpha}{1+\alpha^2}ln\left[ \frac{(\mathrm{Re} R)^2-m^2-l^2+{a^2\over m^2+a^2+l^2}(\mathrm{Im} R)^2}
{(\mathrm{Re} R+m)^2+\{ {a\over \sqrt{m^2+a^2+l^2}}(\mathrm{Im} R)+l \}^2 }\right]. \label{eq26}
\end{eqnarray}
In the case of $l=0$, this solution corresponds to the S-dihole solution obtained 
from the double Wick rotation of black diholes, discussed by Jones et al. \cite{Jones:2004rg}. Our solution (\ref{eq18})-(\ref{eq26}) is the extension of their solution in the dependence on the NUT-parameter. 
\subsection{S-dihole II}
Our second solution is constructed by the following double Wick rotation: 
\begin{equation}
t \to iy \ ,\hspace{.3cm} z\to i\tau \hspace{.3cm}m\to im.
\end{equation}
In this case, the quantity $R_+ + R_- +2m$ appearing in (\ref{fweyl}) is not real.
Then the coordinate $r=\frac{1}{2}(R_+ + R_- +2m)$ does not make sense. 
We can avoid this problem by replacing $R_-\to-R_-$ in the solution(\ref{weyl1})-(\ref{weyl2}); 
this replacement is a choice of branch which appears from the square roots of Weyl functions. 
The metric is represented as
\begin{equation}
ds^2 = \frac{1}{f} \left(\rho^2 d\varphi ^2 + e^{2\gamma}(d \rho^2 -d \tau^2) \right) +f dy^2, \label{eqe18}
\end{equation}
where 
\begin{eqnarray}
R&=& \sqrt{\rho^2-(\tau +i \sqrt{a^2-m^2-l^2})^2}, \\
f&=&\left[ \frac{(\mathrm{Im} R)^2-m^2-l^2+{a^2\over a^2-m^2-l^2}(\mathrm{Re} R)^2}
{(\mathrm{Im} R-m)^2+\{ {a\over \sqrt{a^2-m^2-l^2}}(\mathrm{Re} R)+l \}^2 }\right]^\frac{2}{1+\alpha^2}, \\
e^{2\gamma}&=&\left[ {(\mathrm{Im} R)^2-m^2-l^2+{a^2\over
a^2-m^2-l^2}(\mathrm{Re} R)^2
\over |R|^2}\right]^\frac{4}{1+\alpha^2},
\end{eqnarray}
and
\begin{eqnarray}
A&=&\frac{-2}{\sqrt{1+\alpha^2}\left[ (\mathrm{Im} R)^2-m^2-l^2+\frac{a^2}{a^2-m^2-l^2}(\mathrm{Re} R)^2 \right]}\nonumber \\
& &\times \Big[ \left\{ -am(\mathrm{Im} R-m)+al^2 \right\} (1-\frac{(\mathrm{Re} R)^2}{a^2-m^2-l^2}) 
+ \frac{l(\mathrm{Re} R)\left\{(\mathrm{Im} R)^2+(a^2-m^2-l^2)\right\}}{\sqrt{a^2-m^2-l^2}} \Big] 
d \varphi , \hspace*{.6cm}~~ 
\end{eqnarray}
\begin{eqnarray}
\phi=-\frac{\alpha}{1+\alpha^2}ln\left[ \frac{(\mathrm{Im} R)^2-m^2-l^2+{a^2\over a^2-m^2-l^2}(\mathrm{Re} R)^2}
{(\mathrm{Im} R-m)^2+\{ {a\over \sqrt{a^2-m^2-l^2}}(\mathrm{Re} R)+l \}^2 }\right]. \label{eqe26}
\end{eqnarray}
The difference between S-dihole I (\ref{eq18})-(\ref{eq26}) and S-dihole II (\ref{eqe18})-(\ref{eqe26}) is in the spacetime structure 
\cite{Jones:2004pz}. 
Sending mass parameter $m \to im$ and NUT-parameter $l \to il$ for S-dihole I solution 
produces a S-dihole II. 
To clarify the difference from the work of Jones et al. \cite{Jones:2004rg}, 
let us consider the behaviors of our S-dihole I solution (\ref{eq18})-(\ref{eq26}) in more detail. 

\section{Numerical results and S-charge}
The behaviors of solution (\ref{eq18})-(\ref{eq26}) are shown in Figure 1 for typical values of $a$, $m$, $l$ and $\alpha$. 
The S-dihole I solution is asymptotically flat at large radius, or in the far past or future. 
The metric components are smooth and non-singular for real values of $\rho$ and $t$. 
Outside the light cone $\rho^2=t^2$ in the time-like direction, the fields are connected to the Minkowski space. 
The shapes of gravitational potential $g_{tt}$ and the gauge field $A_{\varphi}$ crossing the light cone represent
that an observer feels gravitational and electromagnetic forces. 
These are typical properties of S-brane solutions. 
For the dependence on NUT-parameter, 
the forces felt by the observer crossing the light cone are the difference between future and past. 
This is difference from the S-brane solution of Jones et al. \cite{Jones:2004rg} : 
in the case of their S-brane solution, the observer feels the same forces. 

\begin{figure}[htpb]
\begin{center}
\vspace{1cm}
\hspace{-1cm}$f$\hspace{7.5cm}$g_{tt}$\\
{\includegraphics[width=6.5cm]{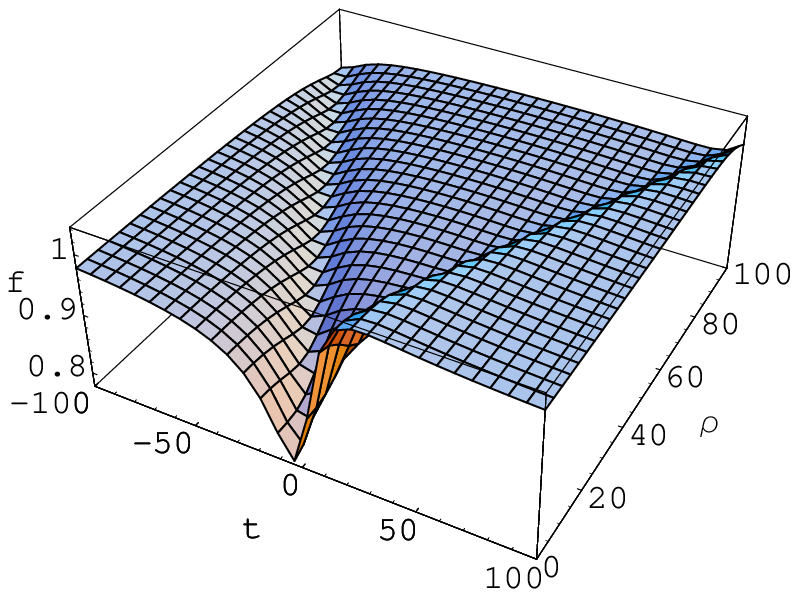}\hspace{1cm}
\includegraphics[width=6.5cm]{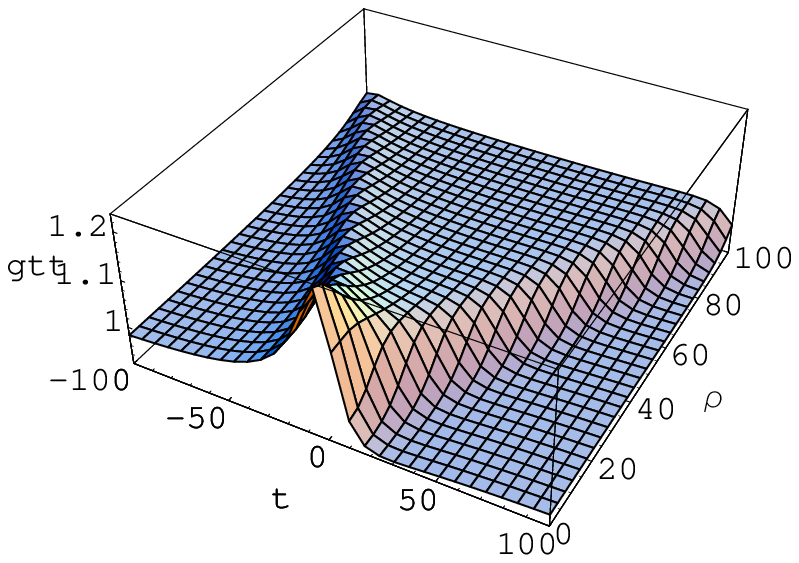}}\\
\hspace{-1cm}$A_{\varphi}$\hspace{7.5cm}$\phi$\\
{\includegraphics[width=6.5cm]{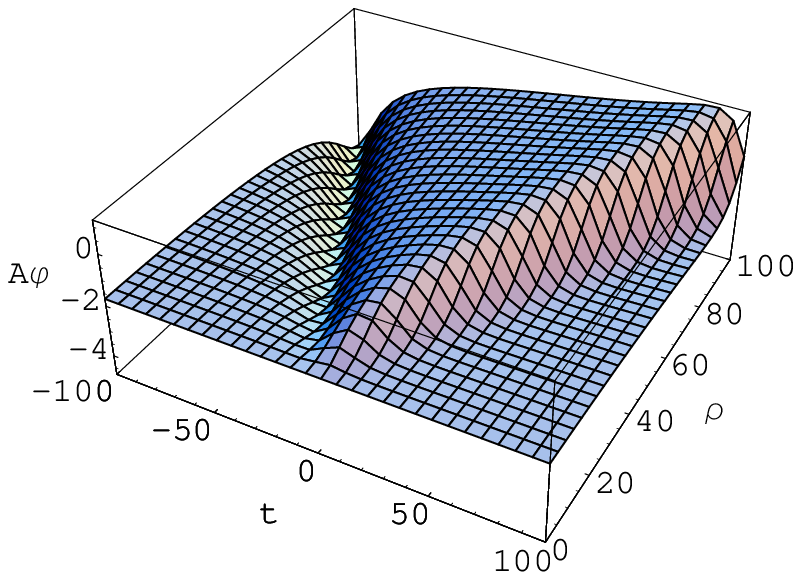}\hspace{1cm}
\includegraphics[width=6.5cm]{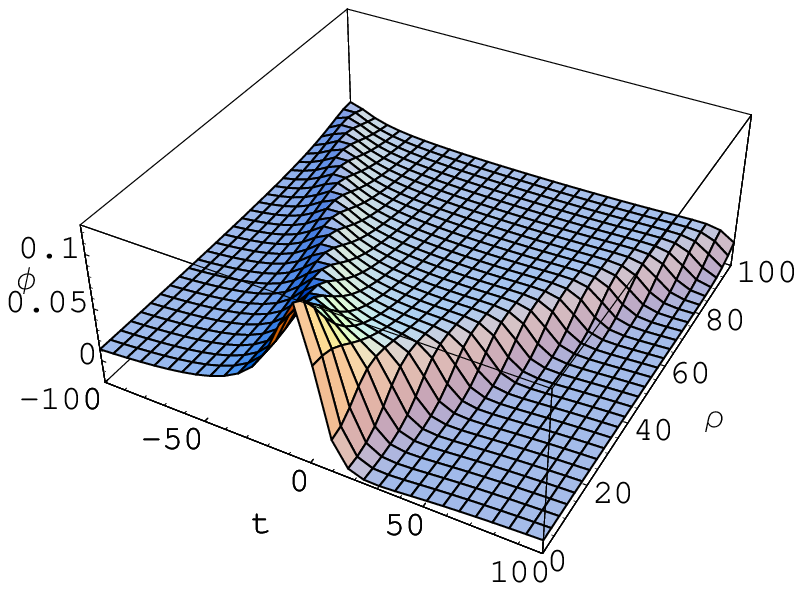}}
\caption{The behaviors of S-dihole I solution for typical values of the parameter.}
\end{center}
\label{ff1p}
\end{figure}
We compute the s-charge of S-dihole I in Weyl coordinates at a constant $\tau$ slice. 
\begin{eqnarray}
Q_s&=&\frac{1}{4\pi} \int  \partial_{\rho} A~d\rho~d\varphi \nonumber \\
&=&\frac{1}{\sqrt{1+\alpha^2}}\left( \frac{m}{a} \sqrt{m^2+a^2+l^2}+\frac{m^2}{a}+\frac{l^2}{a} \right).
\label{Qs}
\end{eqnarray}
This s-charge is conserved at constant $\tau$ slice. 
In the case of $l=0$, (\ref{Qs}) is equal to the s-charge obtained from the black diholes with balanced magnetic charge \cite{Jones:2004rg}. 
Both the behavior of metric components and the conserved s-charge
indicates that our solution obtained from unbalanced black diholes is a kind of the S-brane solutions. 
\section{Conclusion}
We obtained the exact, non-singular and time-dependent solutions. 
Our solutions had the typical properties of S-brane solution 
and the s-charge was conserved over constant time slice. 
It was found that our solution represented one of the S-brane solutions, constructed from the black diholes with the unbalanced magnetic charges. 
The NUT-parameter $l$ represents the net magnetic charge of black diholes. 
In the case of no dilaton and $l=0$, our solution coincides with the S-brane solution discussed in \cite{Jones:2004rg}, where the S-brane solutions constructed form the black diholes. 

Considering the extension of S-dihole solutions, 
Jones et al. \cite{Jones:2004rg} studied the S-brane solution constructed by double 
Wick rotation for array of alternating-charge Reissner-Nodstr\"{o}m black holes. 
Their S-brane solutions are periodic in imaginary time and large-N duals of unstable D-brane creation/decay in string theory. 
Our S-branes are constructed from the black diholes with the unbalanced magnetic charge. 
In a further works, we will study the array of unstable black diholes. 
Using this array solution, we expect that we can study 
the S-brane with the unbalanced magnetic charge and 
the process of more general D-brane decay and creation. 




\begin{thebibliography}{999}
\bibitem{Papapetrou:1947ib}
A.~Papapetrou,
Proc.\ Roy.\ Irish Acad.\ (Sect.\ A)A {\bf 51}, 191 (1947).

\bibitem{Majumdar:eu}
S.~D.~Majumdar,
Phys.\ Rev.\  {\bf 72}, 390 (1947).

\bibitem{Myers:rx}
R.~C.~Myers,
Phys.\ Rev.\ D {\bf 35}, 455 (1987).

\bibitem{IsraelKhan}
W.~Israel and K.~A.~Khan,
Nuovo\ Cim.,\  {\bf 33},\ 331 (1964).

\bibitem{KS3}
K.~Shiraishi,
J.\ Math.\ Phys.\  {\bf 34}, 1480 (1993).

\bibitem{Kastor:1992nn}
D.~Kastor and J.~H.~Traschen,
D {\bf 47}, 5370 (1993)

\bibitem{Tan:2003jz}
H.~S.~Tan and E.~Teo,
Phys.\ Rev.\ D {\bf 68}, 044021 (2003)

\bibitem{Emparan:1999au}
R.~Emparan,
Phys.\ Rev.\ D {\bf 61}, 104009 (2000)

\bibitem{Emparan:2001bb}
R.~Emparan and E.~Teo,
Nucl.\ Phys.\ B {\bf 610}, 190 (2001)

\bibitem{Teo:2003ug} 
E.~Teo,
Phys.\ Rev.\ D {\bf 68}, 084003 (2003)

\bibitem{Costa:2000kf}
M.~S.~Costa and M.~J.~Perry,
Nucl.\ Phys.\ B {\bf 591}, 469 (2000)

\bibitem{Gutperle:2002ai}
M.~Gutperle and A.~Strominger,
JHEP {\bf 0204}, 018 (2002)

\bibitem{Chen:2002yq}
C.~M.~Chen, D.~V.~Gal'tsov and M.~Gutperle,
Phys.\ Rev.\ D {\bf 66}, 024043 (2002)

\bibitem{Kruczenski:2002ap}
M.~Kruczenski, R.~C.~Myers and A.~W.~Peet,
JHEP {\bf 0205}, 039 (2002)

\bibitem{Ohta:2003uw}
N.~Ohta,
Phys.\ Lett.\ B {\bf 558}, 213 (2003)


\bibitem{Burgess:2002vu}
  C.~P.~Burgess, F.~Quevedo, S.~J.~Rey, G.~Tasinato and I.~Zavala,
  JHEP {\bf 0210}, 028 (2002)

\bibitem{Tasinato:2004dy}
  G.~Tasinato, I.~Zavala, C.~P.~Burgess and F.~Quevedo,
  JHEP {\bf 0404}, 038 (2004)

\bibitem{Astefanesei:2005eq}
  D.~Astefanesei and G.~C.~Jones,
  JHEP {\bf 0506}, 037 (2005)

\bibitem{Jones:2004pz}
G.~Jones and J.~E.~Wang,
arXiv:hep-th/0409070.

\bibitem{Jones:2004rg}
G.~Jones, A.~Maloney and A.~Strominger,
Phys.\ Rev.\ D {\bf 69}, 126008 (2004)

\bibitem{Biswas:2004zc}
A.~Biswas,
Phys.\ Lett.\ B {\bf 600}, 157 (2004)

\bibitem{Jones:2005hj}
G.~C.~Jones and J.~E.~Wang,
Phys.\ Rev.\ D {\bf 71}, 124019 (2005).


\bibitem{Liang:2001sp}
Y.~C.~Liang and E.~Teo,
Phys.\ Rev.\ D {\bf 64}, 024019 (2001)

\bibitem{Gibbons:1987ps}
G.~W.~Gibbons and K.~i.~Maeda,
Nucl.\ Phys.\ B {\bf 298}, 741 (1988).

\bibitem{Wang:2002fd}
  J.~E.~Wang,
  JHEP {\bf 0210}, 037 (2002)

\end{thebibliography}
\end{document}